\documentclass[prd,onecolumn,nofootinbib,showpacs,showkeys]{revtex4}
\usepackage{bm,amsmath,amssymb}
\newcommand\gothfamily{\usefont{U}{ygoth}{m}{n}}
\DeclareTextFontCommand{\textgoth}{\gothfamily}

\begin{document}

\title{Mass of the universe in a black hole}
\author{Nikodem J. Pop{\l}awski}
\affiliation{Department of Physics, Indiana University, Swain Hall West, 727 East Third Street, Bloomington, Indiana 47405, USA}
\email{nipoplaw@indiana.edu}
\date{\today}

\begin{abstract}
If spacetime torsion couples to the intrinsic spin of matter according to the Einstein-Cartan-Sciama-Kibble theory of gravity, then the resulting gravitational repulsion at supranuclear densities prevents the formation of singularities in black holes.
Consequently, the interior of every black hole becomes a new universe that expands from a nonsingular bounce.
We consider gravitational collapse of fermionic spin-fluid matter with the stiff equation of state in a stellar black hole.
Such a collapse increases the mass of the matter, which occurs through the Parker-Zel'dovich-Starobinskii quantum particle production in strong, anisotropic gravitational fields.
The subsequent pair annihilation changes the stiff matter into an ultrarelativistic fluid.
We show that the universe in a black hole of mass $M_\textrm{BH}$ at the bounce has a mass $M_\textrm{b}\sim M^2_\textrm{BH} m^{1/2}_\textrm{n}/m^{3/2}_\textrm{Pl}$, where $m_\textrm{n}$ is the mass of a neutron and $m_\textrm{Pl}$ is the reduced Planck mass.
For a typical stellar black hole, $M_\textrm{b}$ is about $10^{32}$ solar masses, which is $10^6$ larger than the mass of our Universe.
As the relativistic black-hole universe expands, its mass decreases until the universe becomes dominated by nonrelativistic heavy particles.
\end{abstract}

\maketitle

The Einstein-Cartan-Sciama-Kibble (ECSK) theory of gravity which naturally extends the general theory of relativity (GR) to account for the quantum-mechanical, intrinsic angular momentum (spin) of elementary particles that compose gravitating matter \cite{KS,Hehl,Tra,Lo}.
This theory is based on the Lagrangian density for the gravitational field that is proportional to the curvature scalar, as in GR \cite{LL}.
It removes, however, the constraint of GR that the torsion tensor (the antisymmetric part of the affine connection) be zero by promoting this tensor to a dynamical variable like the metric tensor \cite{KS,Hehl,Tra,Lo}.
The torsion tensor is then given by the principle of stationary action and in many physical situations it vanishes.
But in the presence of fermions, which compose all stars in the Universe, spacetime torsion does not vanish because Dirac fields couple minimally to the torsion tensor through the affine connection \cite{KS,Hehl,Lo}.
At macroscopic scales, such particles can be averaged and described as a spin fluid \cite{spin_fluid}.
The spin-fluid form of the spin tensor results from the conservation law for this tensor \cite{Hehl,Lo} using the Papapetrou method of multipole expansion \cite{NSH,non}.

The Einstein-Cartan field equations of the ECSK gravity can be written as the general-relativistic Einstein equations with the modified energy-momentum tensor \cite{KS,Hehl}.
Such a tensor has terms which are quadratic in the spin tensor and thus do not vanish after averaging \cite{avert_avg,Kuch}.
These terms are proportional to the square of the fermion number density \cite{spin} and are significant only at densities of matter that are much larger than the nuclear density.
They generate gravitational repulsion in spin-fluid fermionic matter \cite{avert_avg,avert,Kuch}.
Such a repulsion becomes significant at extremely high densities that existed in the early Universe and exist inside black holes.
It prevents a cosmological singularity, replacing it by a state of minimum but finite radius.
This extremely hot and dense state is a big bounce that follows the contracting phase of the Universe and initiates its rapid expansion \cite{Kuch,infl}.
Torsion therefore provides a natural way to explain what caused such an expansion and what existed before the Universe began to expand.

The dynamics of the closed Universe immediately after the big bounce predicts that our Universe appears spatially flat, homogeneous and isotropic because of an extremely small and negative torsion density parameter \cite{infl}.
Thus the ECKS gravity not only eliminates the problem of the initial singularity but also provides a compelling alternative to cosmic inflation.
Although other models with a bounce instead of the big bang also solve the flatness and horizon problems \cite{bounce}, the ECSK gravity is the most natural and simplest theory that solves these problems because it does not introduce additional matter fields or free parameters.

The coupling between spin and torsion may be the mechanism that allows for a scenario in which every black hole produces a new universe inside, instead of a singularity.
The contraction of our Universe before the bounce at the minimum radius may thus correspond to the dynamics of matter inside a collapsing black hole existing in another universe \cite{BH,ER}.
A scenario in which the Universe was born in a black hole seems more reasonable than the contraction of the Universe from infinity in the past because the latter does not explain what caused such a contraction.
This scenario also explains the time asymmetry in a new universe forming inside a black hole because the motion of the collapsing matter through the black hole's event horizon can only happen in one direction and thus it is asymmetric \cite{infl}.
The arrow of time in a daughter universe would thus be inherited, through torsion, from a mother universe.

Torsion appears as a plausible physical phenomenon that may solve some major puzzles regarding our understanding of elementary particles, black holes and the Universe.
It may introduce an effective ultraviolet cutoff in quantum field theory for fermions \cite{non}.
Moreover, torsion modifies the classical Dirac equation by generating the cubic Hehl-Datta term \cite{Dirac} which may be the source of the observed matter-antimatter imbalance and dark matter in the Universe \cite{bar}.
Torsion can also generate massive vectors that are characteristic to electroweak interactions \cite{mass}.
Finally, the gravitational interaction of condensing fermions due to the Hehl-Datta term may be the source of the observed small, positive cosmological constant which is the simplest explanation for dark energy that accelerates the present Universe \cite{dark}.

Observations of thermal emission during X-ray bursts suggest that the matter inside neutron stars is characterized by a stiff equation of state $p=\epsilon$ \cite{stiff_NS}, where $p$ is the pressure and $\epsilon$ is the energy density.
Theoretical models of ultradense matter predict the stiff equation of state at very high densities from the strong interaction of the nucleon gas \cite{stiff_mat}.
We can therefore expect that the matter collapsing inside stellar black holes is stiff too.
Stiff matter has also been suggested as a possible content of the Universe during early stages of its expansion \cite{stiff_cosm,stiff_BH}.
Furthermore, a black hole grows as fast as the Universe if the equation of state is stiff \cite{stiff_BH}, showing that the black hole's interior composed of stiff matter is dynamically equivalent to a universe.
Nonnucleonic degrees of freedom at supranuclear densities can soften the equation of state \cite{Lat}.
Such a softening would decrease the mass of the Universe inside a black hole.
The Shapiro-delay measurement of the mass of the neutron star in a binary pulsar puts constraints on the equation of state of the neutron star \cite{Dem}.
The data from the PSR J1614-2230 pulsar effectively rule out the presence of exotic matter in the form of kaon condensates, hyperons, heavy bosons, or free quarks at supranuclear densities \cite{Dem}.
These data therefore support a stiff equation of state in neutron stars and black holes.

If the matter collapsing in a forming black hole is stiff then the covariant conservation of the modified energy-momentum tensor causes that the mass of the universe inside this black hole increases by many orders of magnitude, as we show below.
Physically, such an increase of mass occurs by the Parker-Zel'dovich-Starobinskii quantum particle production in the presence of strong, anisotropic gravitational fields \cite{part}.
Such a pair production does not change the equation of state of the collapsing matter, which remains stiff because of the ultradense regime.
The subsequent pair annihilation into photons, however, changes the stiff matter into an ultrarelativistic fluid and results in isotropization of the universe in a black hole, which stops the Parker-Zel'dovich-Starobinskii process.
The pair production also does not change the total (matter plus gravitational field) energy of the universe in a black hole, which is equal to zero (even in the presence of a cosmological constant) \cite{energy}.
The formation and evolution of such a universe (including the above increase of its mass) is not visible for observers outside the black hole, for whom the event horizon's formation and all subsequent processes would occur after an infinite amount of time had elapsed \cite{LL}.
Gravitational time dilation thus makes it possible for the mass of the Universe to be much bigger than the mass of the parent black hole as measured by external observers.

The Einstein-Cartan field equations for the closed Friedman-Lema\^{i}tre-Robertson-Walker (FLRW) metric describing such a universe are given by the Friedman equations for the scale factor $a(t)$ \cite{infl,spin}:
\begin{eqnarray}
& & {\dot{a}}^2+k=\frac{1}{3}\kappa\Bigl(\epsilon-\frac{1}{4}\kappa s^2\Bigr)a^2+\frac{1}{3}\Lambda a^2, \label{Fri1} \\
& & \frac{d}{dt}\bigl((\epsilon-\kappa s^2/4)a^3\bigr)+(p-\kappa s^2/4)\frac{d}{dt}(a^3)=0,
\label{Fri2}
\end{eqnarray}
where dot denotes differentiation with respect to $ct$, $k=1$, $s^2$ is the square of the spin tensor \cite{avert_avg}, and $\Lambda$ is the cosmological constant.
For a fluid consisting of unpolarized fermions, this quantity is given by
\begin{equation}
s^2=\frac{1}{8}(\hbar cn)^2,
\label{spide}
\end{equation}
where $n$ is the fermion number density \cite{spin}.
Equation (\ref{Fri2}) corresponds to the covariant conservation of the modified energy-momentum tensor.
For the stiff equation of state and for densities at which $\kappa s^2\ll\epsilon$, this equation gives $\epsilon\propto a^{-6}$, which then gives $E\propto a^{-3}$, where $E$ is the total energy of matter in the Universe.
Thus the total mass of matter in the Universe $m$ scales according to $m\propto a^{-3}$, which gives $N\propto a^{-3}$, where $N$ is the number of fermions in the Universe (we assume that this number is proportional to the total number of particles).
Accordingly, $n\propto a^{-6}$, from which we obtain $s^2\propto a^{-12}$ because of (\ref{spide})
Substituting (\ref{spide}) into the Friedman equation (\ref{Fri1}) gives then \cite{massUni}
\begin{equation}
{\dot{a}}^2+k=\frac{1}{3}\kappa\epsilon_0\frac{a_0^6}{a^4}-\frac{1}{96}(\hbar c\kappa)^2 n_0^2\frac{a_0^{12}}{a^{10}},
\label{master}
\end{equation}
where $\epsilon_0$ and $n_0$ correspond to the universe at $a=a_0$.
We neglect the last term in (\ref{Fri1}) with the cosmological constant because it is negligibly small for $a<a_0$.

A universe born in a black hole begins to contract when $\dot{a}=0$ at $a_0=r_g$, where $r_g=2GM_\textrm{BH}/c^2$ is the Schwarzschild radius of the black hole of mass $M_\textrm{BH}$.
At this stage, the second term on the right-hand side of (\ref{master}) is negligible, yielding $\epsilon_0=3M_\textrm{BH}c^2/(4\pi r_g^3)$, which coincides with the rest energy density of the black hole regarded as a uniform sphere \cite{massUni}.
If the matter in a black hole at $a=a_0$ is composed of neutrons, which is a reasonable assumption, then the initial number density of fermions in this universe is $n_0=9M_\textrm{BH}/(4\pi r_g^3 m_\textrm{n})$, where $m_\textrm{n}$ is the mass of a neutron \cite{massUni}.
The universe in the black hole contracts until the negative second term on the right-hand side of (\ref{master}) becomes significant and counters the first term ($k$ is negligible in this regime).
At this moment, where $a=a_\star$, the condition $\dot{a}=0$ in (\ref{master}) gives $a_\star=(\kappa/(32\epsilon_0))^{1/6}(\hbar cn_0)^{1/3}a_0$, which leads to the scale factor of the universe at the bounce \cite{massUni}:
\begin{equation}
a_\star=\Bigl(\frac{27}{128}\Bigr)^{\frac{1}{6}}\frac{2G}{c^2}\Bigl(\frac{M_\textrm{BH}^2 m_\textrm{Pl}^2}{m_\textrm{n}}\Bigr)^{\frac{1}{3}},
\label{astar}
\end{equation}
where $m_\textrm{Pl}=(\hbar/(\kappa c^3))^{1/2}$ is the reduced Planck mass.
The mass of the universe at the bounce is given by $m_\star=(a_0/a_\star)^3 M_\textrm{BH}$ (since $m\propto a^{-3}$).
Substituting (\ref{astar}) into this relation gives \cite{massUni}
\begin{equation}
m_\star=\Bigl(\frac{128}{27}\Bigr)^{\frac{1}{2}}\frac{M_\textrm{BH}^2 m_\textrm{n}}{m_\textrm{Pl}^2}.
\label{mstar}
\end{equation}
After the bounce, the ultrarelativistic matter in a black hole expands as a new, closed universe \cite{infl,massUni}.
As the black-hole universe expands, its mass decreases until the universe becomes dominated by nonrelativistic particles.
If the mass of the resulting nonrelativistic universe exceeds a critical value $m_c\sim c^2/(G\sqrt{\Lambda})$, then the universe expands to infinity \cite{Lo,massUni,dyn}.
As time tends to infinity, the parent black hole becomes an Einstein-Rosen bridge that connects the daughter universe in the black hole to the mother universe.
The spin-torsion coupling in the ECSK gravity therefore allows Nature to choose a mathematical solution of the gravitational field equations that is regular (an Einstein-Rosen bridge) \cite{ER} rather than a mathematical solution that is singular (a Schwarzschild black hole).
In the daughter universe, the mother universe appears as the only white hole in space.

The mass density at the bounce is equal to $\rho_\star=m_\star/(4\pi a^3_\star)$, which gives
\begin{equation}
\rho_\star=\frac{128}{27}(8\pi)^2\rho_\textrm{C},\,\,\,\rho_\textrm{C}=\frac{m^2_\textrm{n}c^4}{G\hbar^2},
\label{rhostar}
\end{equation}
where $\rho_\textrm{C}$ is the Cartan density for neutrons \cite{Tra,non}.
The quantity $\rho_\star$ is independent of the mass of the black hole and its order of the Cartan density indicates that the neutrons are nonrelativistic.
The above analysis \cite{massUni}, however, does not include the annihilation of the pairs produced through the Parker-Zel'dovich-Starobinskii process into radiation.
Let as assume, for simplicity, that these pairs start annihilating to radiation when $a=a_\star$.
As a result, the fermion number density decreases, weakening the gravitational repulsion from the spin-torsion coupling.
The gravitational collapse can therefore proceed further, until the spin-torsion term $\kappa s^2/4$ in the Friedman equation (\ref{Fri1}) balances the relativistic energy density $\epsilon$.
If the ultrarelativistic matter contains only the known standard-model particles (with equal temperatures), then such a balance (big bounce) occurs at $\epsilon_\textrm{b}=15.4\,\epsilon_\textrm{Pl}$, where $\epsilon_\textrm{Pl}=\rho_\textrm{Pl}c^2$ and $\rho_\textrm{Pl}=c^5/(\hbar G^2)$ is the Planck density \cite{bb}.
During the ultrarelativistic collapse, $\epsilon\propto a^{-4}$, which gives $m\propto a^{-1}$.
The bounce thus occurs not at $a=a_\star$, but at $a=a_\textrm{b}=a_\star(\rho_\star c^2/\epsilon_\textrm{b})^{1/4}$, which gives
\begin{equation}
a_\textrm{b}=a_\star\biggl(\frac{m_\textrm{n}}{m_\textrm{Pl}}\biggr)^{\frac{1}{2}}\biggl(\frac{128}{27}\frac{(8\pi)^2}{15.4}\biggr)^{\frac{1}{4}}=\frac{2G}{c^2}\bigl(M_\textrm{BH}^{2/3}m_\textrm{Pl}^{1/6}m_\textrm{n}^{1/6}\bigr)\Bigl(\frac{128}{27}\Bigr)^{\frac{1}{12}}\frac{(8\pi)^{1/2}}{15.4^{1/4}}.
\label{ab}
\end{equation}
Accordingly, the mass of the universe in a stellar black hole at the bounce is not $m_\star$, but is equal to $M_\textrm{b}=m_\star a_\star/a_\textrm{b}$, which gives
\begin{equation}
M_\textrm{b}=m_\star\biggl(\frac{m_\textrm{Pl}}{m_\textrm{n}}\biggr)^{\frac{1}{2}}\biggl(\frac{27}{128}\frac{15.4}{(8\pi)^2}\biggr)^{\frac{1}{4}}=M^2_\textrm{BH} m^{-3/2}_\textrm{Pl} m^{1/2}_\textrm{n}\biggl(\frac{128}{27}\frac{15.4}{(8\pi)^2}\biggr)^{\frac{1}{4}}.
\label{mb}
\end{equation}
The values (\ref{ab}) and (\ref{mb}) refine the results of \cite{massUni}.

For a typical stellar black hole, $M_\textrm{b}$ is about $10^{32}$ solar masses, which is $10^6$ larger than the mass of our Universe \cite{infl,massUni}.
As the ultrarelativistic black-hole universe expands, however, its mass decreases until the universe becomes dominated by nonrelativistic particles.
Such particles may be heavy fermions that exist at very high energies and are postulated to be the source of baryogenesis and dark matter (\cite{bar} and references therein).
In this case, it could be possible that the mass of the universe in a black hole decreases only by a desired factor $10^6$, asymptotically matching the mass of our Universe.
The value (\ref{mb}) can also decrease if we consider a more realistic scenario in which the Parker-Zel'dovich-Starobinskii particle-antiparticle pairs start annihilating to radiation earlier than at $a=a_\star$.
Such a scenario, as well as the dynamics of the pair production, will be investigated further.

\end{document}